\documentstyle[12pt]{article} 
\newfont{\Mb}{msbm10}

\begin{document}
\setcounter{equation}{0}
\setcounter{figure}{0}
\setcounter{table}{0}

\hspace\parindent
\thispagestyle{empty}

\bigskip
\bigskip
\bigskip
\begin{center}
{\LARGE \bf A semi-algorithm to find elementary first order invariants of rational
second order ordinary differential equations}
\end{center}

\bigskip

\begin{center}
{\large
$^{a,b}$J. Avellar, $^a$L.G.S. Duarte, $^{c}$ S.E.S. Duarte and $^a$L.A.C.P. da Mota \footnote{E-mails: javellar@dft.if.uerj.br, lduarte@dft.if.uerj.br, sduarte@dft.if.uerj.br and damota@dft.if.uerj.br}
}

\end{center}

\bigskip
\centerline{\it $^a$ Universidade do Estado do Rio de Janeiro,}
\centerline{\it Instituto de F\'{\i}sica, Depto. de F\'{\i}sica Te\'orica,}
\centerline{\it 20559-900 Rio de Janeiro -- RJ, Brazil}

\bigskip
\centerline{\it $^b$ Funda\c c\~ao de Apoio \`a Escola T\'ecnica,}
\centerline{\it E.T.E. Juscelino Kubitschek,}
\centerline{\it 21311-280 Rio de Janeiro -- RJ, Brazil}

\bigskip
\centerline{\it $^c$ Funda\c c\~ao de Apoio \`a Escola T\'ecnica,}
\centerline{\it E.T.E. Visconde de Mau\'a,}
\centerline{\it 20537-200 Rio de Janeiro -- RJ, Brazil}

\bigskip
\bigskip

\abstract{ Here we present a method to find elementary first
integrals of rational second order ordinary differential equations
(SOODEs) based on a Darboux type procedure
\cite{ManMac,firsTHEOps1,secondTHEOps1}. Apart from practical
computational considerations, the method will be capable of
telling us (up to a certain polynomial degree) if the SOODE has an
elementary first integral and, in positive case, finds it via
quadratures.}

\bigskip
\bigskip
\bigskip
\bigskip
\bigskip
\bigskip




\newpage

\section{Introduction}
\label{intro}

The differential equations (DEs) are the most widespread way to
formulate the evolution of any given system in many scientific areas.
Therefore, for the last three centuries, much effort has been made in
trying to solve them.

Broadly speaking, we may divide the approaches to solving ODEs in
the ones that classify the ODE and the ones that do not
(classificatory and non-classificatory methods).  Up to the end of
the nineteenth century, we only had many (unconnected)
classificatory methods to try to deal with the solving of ODEs.
Sophus Lie then introduced his method ~\cite{step,bluman,olver}
that was meant to be general and try to solve any ODE, i.e.,
non-classificatory. Despite this appeal, the Lie approach had a
shortcoming: namely, in order to deal with the ODE, one has to
know the symmetries of the given ODE. Unfortunately, this part of
the procedure was not algorithmic (mind you that the
classificatory approach is algorithmic by nature). So, for many
decades, the Lie method was not put to much ``practical'' use
since to ``guess'' the symmetries was considered to be as hard as
guessing the solution to the ODE itself. In \cite{nossolie1,
nossolie2}, an attempt was made to make this searching for the
symmetries to the ODE practical and, consequently, make the Lie
method more used.

Even thought the attempt mentioned above was very successful, the
procedures applied to find the symmetries were heuristic. So, a
non- classificatory {\bf algorithmic} approach was still missing.
The first algorithmic approach applicable to solving first order
ordinary differential equations (FOODEs) was made by M. Prelle and
M. Singer \cite{PS}. The attractiveness of the PS method lies not
only in its basis on a totally different theoretical point of view
but, also in the fact that, if the given FOODE has a solution in
terms of elementary functions, the method guarantees that this
solution will be found (though, in principle it can admittedly
take an infinite amount of time to do so). The original PS method
was built around a system of two autonomous FOODEs of the form
$\dot{x} = P(x,y)$, $\dot{y}={\cal P}(x,y)$ with $P$ and ${\cal
P}$ in ${\it C}[x,y]$ or, equivalently, the form $y'=R(x,y)$, with
$R(x,y)$ a rational function of its arguments.

The PS approach has its limitations, for instance, it deals only
with {\bf rational} FOODEs. But, since it is so powerful in many
respects, it has generated many extensions
\cite{Shtokhamer,collins,chris1,chris2,llibre,firsTHEOps1,secondTHEOps1,nossoPS1CPC}

Nevertheless, all these extensions deal only with FOODEs. In particular,
the second order ordinary differential equations (SOODEs) play a very
important role, for instance, in the physical sciences. So, with this in
mind, we have produced \cite{PS2} a PS-type approach to deal with
SOODEs. This approach dealt with SOODEs that presented
elementary\footnote{For a formal definition of elementary function, see
\cite{davenport}.} solutions (with
two elementary first order invariants).

Here, we present a different approach that, besides dealing with a much
broader class of SOODEs (those with at least one elementary first order
invariant), does not depend on a conjecture about the general structure
of the first order invariants.

In section \ref{earlier}, we present the state of the art up to
the present paper. In the following section, we introduce some
important theoretical results for the building of the algorithm to
find the integrating factor. In section \ref{findingR}, we present
the algorithm for finding the integrating factor with examples of
its application. Finally, we present our conclusions and point out
some directions to further our work.

\section{Earlier Results}
\label{earlier}

In the paper \cite{PS}, one can find an important result that, translated to
the case of SOODEs of the form
\begin{equation}
\label{soode}
y'' = {\frac{M(x,y,y')}{N(x,y,y')}} = \phi(x,y,y'),
\end{equation}
where $M$ and $N$ are polynomials in $(x,y,y')$, can be stated as:
\bigskip

{\bf Theorem 1: }{\it If the SOODE (\ref{soode}) has a first order invariant that can be written
in terms of elementary functions, then it has one of the form:
\begin{equation}
\label{eq_I}
I = w_0 + \sum_i^m c_i \ln (w_i),
\end{equation}
where $m$ is an integer and the $w's$ are algebraic functions\footnote{For a formal definition of
algebraic function, see \cite{davenport}.} of $(x,y,y')$}.
\bigskip

The integrating factor for a SOODE of the form (\ref{soode}) is defined by:

\begin{equation}
\label{Rdefinition}
R (\phi-y'') = \frac{dI(x,y,y')}{dx}
\end{equation}
where $\frac{d}{dx}$ represents the total derivative with respect to
$x$.

Bellow we will present some results and definitions (previously
presented on \cite{PS2}) that we will need. First let us remember that, on the
solutions, $dI = I_x\,dx+I_y\,dy+I_{y'}\,dy'= 0$. So, from equation
(\ref{Rdefinition}), we have:

\begin{equation}
\label{dI}
R (\phi\,dx-dy') = I_x\,dx+I_y\,dy+I_{y'}\,dy' = dI = 0.
\end{equation}
Since $y'\,dx=dy$, we have
\begin{equation}
\label{di}
R \left[(\phi+S\,y')\,dx-S\,dy-dy'\right] = dI = 0,
\end{equation}
adding the null term $S\,y'\,dx-S\,dy$, where $S$ is a function
of $(x,y,y')$. From equation (\ref{di}), we have:
\begin{eqnarray}
\label{compatcond}
I_x &=& R (\phi +  S y'),\nonumber \\
I_y &=& - R S, \\
I_{y'} &=& - R, \nonumber
\end{eqnarray}
that  must satisfy the compatibility conditions. Thus,
defining the differential operator $D$:

\begin{equation}
D \equiv \partial_{x} + y' \partial_{y} + \phi\, \partial_{y'},
\end{equation}
after a little algebra, that can be shown to be equivalent to:
\begin{eqnarray}
\label{SR1}
D[R]  & = & -R (S + \phi_{y'}),\\
\label{SR2}
D[RS] & = & -R \phi_{y}.
\end{eqnarray}


\section{New theoretical results concerning the function $S$}
\label{NewTheo}

Let us start this section by stating a corollary to {\bf theorem 1}
concerning $S$ and $R$.

\bigskip
{\bf Corollary 1: }{\it If a SOODE of the form (\ref{soode}) has a
first order elementary invariant then the integrating factor $R$ for such an
SOODE and the function $S$ defined in the previous section can be written as
algebraic functions of $(x,y,y')$.}

\bigskip
\bigskip
{\bf Proof:}
Using the above mentioned result by Prelle and Singer,
there is always a first order invariant $I=w_0 + \sum_i^m c_i \ln (w_i)$
for the SOODE. So we have, using equation (\ref{Rdefinition}),
\begin{equation}
\label{RIz}
R (\frac{M}{N}-y'') = I_x+y'I_y+y''I_{y'}\Rightarrow R = -I_{y'}
\end{equation}
where $I_u\equiv\partial_uI$. From equation (\ref{eq_I}), we have:
\begin{equation}
\label{dIdx}
I_{y'}={w_0}_{y'} + \sum_i^m c_i\frac{{w_i}_{y'}}{w_i}.
\end{equation}
Then $I_{y'}$ is an algebraic function of $(x,y,y')$ and, by equation
(\ref{RIz}), so is $R$.

From equations (\ref{compatcond}), one can see that:

\begin{equation}
S = \frac{I_y}{I_{y'}} = \frac{{w_0}_{y} + \sum_i^m c_i\frac{{w_i}_{y}}{w_i}}{{w_0}_{y'} + \sum_i^m c_i\frac{{w_i}_{y'}}{w_i}}.
\end{equation}
Therefore, $S$ is also an algebraic function of $(x,y,y')$.$\Box$

\bigskip
\bigskip
Besides that, working on equations (\ref{SR1}) and (\ref{SR2}), we get\cite{PS2}:

\begin{equation}
\label{DS}
D[S] = S^2 + \phi_{y'}\,S - \phi_{y}=\frac{{\cal M}}{{\cal N}},
\end{equation}
where ${\cal M}$ and ${\cal N}$ are given by
\begin{eqnarray}
\label{Ms}
{\cal M} & \equiv & (NS)^2+ (NM_{y'}-MN_{y'})\,S-(NM_{y}-MN_{y}),\\
\label{Ns}
{\cal N} & \equiv & N^2.
\end{eqnarray}

Concerning eq.(\ref{DS}) we can demonstrate the following theorem:
\bigskip

{\bf Theorem 2: }{\it Consider the operator defined by $D_S \equiv
{\cal M}\,\partial_S + {\cal N}\,D$. If $P$ is an eigenpolynomial
of $D_S$ (i.e., $D_S[P] = \lambda P$, where $\lambda$ is a
polynomial) that contains $S$, then $P = 0$ defines a particular
solution of eq.(\ref{DS}). Conversely, If $P$ is a polynomial that
contains $S$, such that $P = 0$ defines a particular solution of
eq.(\ref{DS}), then $P$ is either an eigenpolynomial of $D_S$ or
$P$ is an absolute invariant of the Lie transformation group
defined by $D_S$.}
\bigskip
\bigskip

\noindent
{\bf Proof:}
In order to demonstrate theorem 2 we will, first, prove the following lema:
\bigskip

{\bf Lema 1: }{\it If eq.(\ref{DS}) ($D[S] = {\cal M}/{\cal N}$) has an
algebraic solution defined by $\sum_i a_i S^i=0$, where the $a_i$ are
polynomials in $(x,y,y')$, then
\begin{equation}
\label{DA}
\sum_i \left( {\cal N} D[a_i]S^i + {\cal M} a_i i S^{i-1} \right)=0.
\end{equation}

Conversely, if $\,\sum_i \left( {\cal N} D[a_i]S^i + {\cal M} a_i i S^{i-1}
\right)=0$ then $\sum_i a_i S^i=0$ defines an algebraic function ($S$) as a particular
solution of eq.(\ref{DS}).

}
\bigskip

{\bf Proof of Lema 1:} We begin by proving the first part of the Lema. Let $\sum_i a_i S^i=0$ define an algebraic particular
solution of eq.(\ref{DS}). Applying the operator $D$ on $\sum_i a_i S^i=0$, one gets:

$$\,\sum_i \left(D[a_i]S^i + a_i i S^{i-1} D[S]\right)=0 \,\,\,\Rightarrow \,\,\,\sum_i \left(D[a_i]S^i + a_i i S^{i-1} \frac{{\cal M}}{{\cal N}}\right)=0.$$

Multiplying this by ${\cal N}$ we get eq.(\ref{DA}).

Let us now prove the converse. Consider that eq.(\ref{DA}) applies.
Multiplying it by ${\cal N}$ and remembering that $S$ obeys $D[S] = {\cal M}/{\cal N}$,
one gets:
\begin{equation}
\label{DSzero}
\sum_i \left(D[a_i]S^i + a_i i S^{i-1}
\frac{{\cal M}}{{\cal N}}\right)\,=0 \Rightarrow -\,\sum_i
\left(D[a_i]S^i\right)/\sum_i \left( a_i i S^{i-1}
\right)\,=\,\frac{{\cal M}}{{\cal N}}
\end{equation}

However, note that, if we have an algebraic function ($B$) defined by
$\sum_i b_i B^i=0$, then $D[B]$ can be found as follows:

$$ D  \left[ \sum_i b_i B^i \right] \,=0\,\Rightarrow \sum_i \left(D[b_i]B^i + b_i i B^{i-1} D[B]\right)\,=0\,\Rightarrow $$
$$D[B]=-\,\sum_i
\left( D[b_i]B^i\right)/\sum_i \left( b_i i B^{i-1}
\right)$$

Therefore, one can see that eq.(\ref{DSzero}) can be put in the form:
\begin{equation}
D[S] = \frac{{\cal M}}{{\cal N}},
\end{equation}
where $S$ is an algebraic function defined by $\sum_i a_i S^i=0$.$\,\,\Box$

\bigskip
\bigskip

Now, using Lema 1, we are going to demonstrate {\bf Theorem 2}.

Consider that $P$ is an eigenpolynomial of $D_S$ that contains
$S$. By definition, $D_S[P] = \lambda P$, where $\lambda$ is a
polynomial. So, writing $P=\sum_i a_i S^i$, where the $a_i$'s are
polynomials in $(x,y,y')$, we have:
\begin{equation}
\sum_i \left( {\cal N} D[a_i]S^i + {\cal M} a_i i S^{i-1}\right)\,=\,\lambda\,P
\end{equation}

Over the algebraic function defined by $P=0$, we get $\sum_i \left( {\cal N}
D[a_i]S^i + {\cal M} a_i i S^{i-1}\right)\,=\,0.$ This, using Lema 1, implies
that $\sum_i a_i S^i=0$ defines a particular solution to eq.(\ref{DS}).

Conversely, consider that $P$ is a polynomial that contains $S$,
such that $P=\sum_i a_i S^i= 0$ defines a particular solution of
eq.(\ref{DS}). Again, via Lema 1, we have that:
\begin{equation}
\label{DsP}
\sum_i \left( {\cal N} D[a_i]S^i + {\cal M} a_i i S^{i-1}\right)\,=\,0.
\end{equation}

Consider now the following operator:
\begin{equation}
{\cal O} =
N^2(x_1,x_2,x_3)\left(\partial_{x_1}+x_3\partial_{x_2}+\frac{M(x_1,x_2,x_3)}{N(x_1,x_2,x_3)}\partial_{x_3}\right) +
{\cal M}(x_1,x_2,x_3,x_4)\partial_{x_4}
\end{equation}
Applying {\cal O} to a generic polynomial ${\cal P}=\sum_ic_ix_4^i$, where the $c_i$'s are
polynomials in $(x1,x2,x3)$, one obtains:
\begin{equation}
\label{OQ}
{\cal O}[{\cal P}]\,=\,\sum_i \left( N^2(x_1,x_2,x_3) {\cal D}[c_i]x_4^i + {\cal M}(x_1,x_2,x_3,x_4) c_i i x_4^{i-1}\right)
\end{equation}
where ${\cal D} \equiv
\left(\partial_{x_1}+x_3\partial_{x_2}+\frac{M(x_1,x_2,x_3)}{N(x_1,x_2,x_3)}\partial_{x_3}\right)$.

\bigskip
Since the terms multiplying the partial derivatives are polynomials,
applying ${\cal O}$ to a polynomial will generate a polynomial. So, from
eq.(\ref{OQ}), we have:
\begin{equation}
\label{OQ2}
\sum_i \left( N^2(x_1,x_2,x_3) {\cal D}[c_i]x_4^i + {\cal M}(x_1,x_2,x_3,x_4)
c_i i x_4^{i-1}\right)\,=\,Q
\end{equation}
where $Q$ is a polynomial in $(x_1,x_2,x_3,x_4)$.

Note that the left-hand side of equations (\ref{OQ2}) and (\ref{DsP})
are formally equivalent. Consider that the hypothesis of the theorem 2
apply, i.e., ${\cal P}=\sum_ic_ix_4^i=0$ defines a algebraic function
$x_4(x_1,x_2,x_3)$ that is a particular solution of
$D[x_4]={\cal M}(x_1,x_2,x_3,x_4)/{\cal N}(x_1,x_2,x_3)$. Lema 1 implies then
that, if we substitute the function $x_4=RootOf(\sum_ic_ix_4^i)$ into
eq.(\ref{OQ2}), we get $0 = Q$. Therefore,
$\,{\cal P}=0\,\,\,\Rightarrow\,\,\,Q=0$. Since both ${\cal P}$ and $Q$
are polynomials we have two possibilities:
\begin{itemize}
\item $Q$ is identically null - in that case, ${\cal P}$ is an
absolute invariant of the Lie transformation group defined by the
operator ${\cal O}$.
\item ${\cal P}$ is a factor of $Q$ - in that case, $Q=\lambda
{\cal P}$, where $\lambda$ is a polynomial in $(x_1,x_2,x_3,x_4)$ and $Q$ is a relative
invariant of the Lie transformation group defined by the operator ${\cal
O}$.
\end{itemize}
This completes the proof of theorem 2.$\Box$

\bigskip
\bigskip
\bigskip
From the results above we may finally conclude:

\bigskip
{\bf Corollary 2:  }{\it If the SOODE (\ref{soode}) has an elementary first
integral, then there is a polynomial $P$ (containing $S$) that is either an
eigenpolynomial of $D_S$ or is an absolute invariant of the Lie
transformation group defined by $D_S$.}

\bigskip
\bigskip

{\bf Proof:} From corollary 1, since eq.(\ref{soode}) has a first
order elementary invariant, there is an algebraic function $S$
that satisfies eq.(\ref{DS}). So, by definition, there is a
polynomial $P$ (that contains $S$) such that $P=0$ defines $S$.
From theorem 2, this implies that the polynomial $P$ is either an
eigenpolynomial of $D_S$ or is an absolute invariant of the Lie
transformation group defined by $D_S$.$\Box$

\bigskip
\bigskip
These theoretical results provide us an algorithm to find $S$. Briefly, in words, what we have
to do is the following:
\begin{itemize}
\item  Find the eigenpolynomials of the $D_S$ operator containing
$S$. Let us call them $P_i$.
\item In order to find $S$, we have to choose one
of those $P_i$ (let us call it $P$) that contains $S$ and solve the equation $P=0$ for $S$.
\item If you succeed in doing the above steps, you would have found $S$ for the SOODE you are dealing with.
\end{itemize}

The importance of these results and the above sketched method is that,
as we shall briefly see, the finding of the algebraic function $S$ will
allow us to produce a semi-algorithmic method to find the integrating
factor (and, consequently, the first order invariant) for SOODEs of the type described in eq.(\ref{soode}).


\section{Finding the integrating factor and the first integral}
\label{findingR}

In what follows, we are going to demonstrate a result concerning the
general structure of $R$, Let us state that result as a theorem.

{\bf Theorem 3:  } {\it Consider a SOODE of the form (\ref{soode}), where $M$
and $N$ are polynomials in $(x,y,y')$, that presents an
elementary first order invariant $I$. Then the integrating factor $R$ for this SOODE
can be written as:
\begin{equation}
\label{Rfi}
R = \prod_i f^{n_i}_i
\end{equation}
where $f_i$ are irreducible polynomials in $(x,y,y',S)$, which are
eigenpolynomials of the operator ${\cal D} \equiv N\,D$ or are factors
of N and $n_i$ are non-zero rational numbers.}

\bigskip
\noindent
{\bf Proof:}

Suppose that the hypothesis of the theorem is satisfied. Re-writing
equation (\ref{dI}):
\begin{equation}
\label{dI2}
\frac{R}{N} \left[(M+S\,N\,y')\,dx-S\,N\,dy-N\,dy'\right] = dI = 0
\end{equation}

For the sake of simplicity, let us establish some notation:
$\overline{M} \equiv (M+y'\,N\,S)$, $\overline{N}\equiv - (N\,S)$
and $\overline{R}\equiv(\frac{R}{N})$. We can write
$I_x=\overline{R}\overline{M}$ and $I_y=-\overline{R}\overline{N}$ and
imposing the compatibility condition $I_{xy}=I_{yx}$, we have
 $\partial_y(\overline{R}\overline{M}) = -
\partial_x(\overline{R}\overline{N})$. Expanding this
\begin{equation}
\label{PSbarra}
\overline{R}_y\,\overline{M}+\overline{R}_x\,\overline{N} =
-\overline{R}(\overline{M}_y+\overline{N}_x)\,\,\,\,\,
\Rightarrow\,\,\,\,\,\frac{\overline{D}\,[\overline{R}]}{\overline{R}}=
-(\overline{M}_y+\overline{N}_x)
\end{equation}
where $\overline{D} \equiv \overline{N} \partial_x + \overline{M} \partial_y$.

Since $\overline{M}$ and $\overline{N}$ are polynomials in
$(x,y,S)$\footnote{Note that $\overline{M}$ and $\overline{N}$ are polynomials in $y'$ as well.}, equation (\ref{PSbarra}) is formally equivalent to the equation
 which establish the conditions for the theorem by Prelle and Singer \cite{PS} with the extension by Shtokhamer
\cite{Shtokhamer}\footnote{Notice that, since $S$ is an algebraic
function of $(x,y,y')$, it is a root of a polynomial equation of the
form $\sum_{i}p_{i}S^{i}=0$, where $p_i$ are polynomials in $(x,y,y')$.
Therefore, its derivatives can be expressed in terms of rational
powers of itself. So, one may consider that would be dealing with a
Shtokhamer extension with $U={S}$.
}. More formally, consider the FOODE defined by:
\begin{equation}
\label{ODEtilde}
y'= \frac{\widetilde{M}}{\widetilde{N}}
\end{equation}
where $\widetilde{M}$ and $\widetilde{N}$ are the polynomials obtained
by replacing $y'$ by a constant $k$ in $\overline{M}$ and $\overline{N}$
respectively.  Then we can write the solution for the FOODE
(\ref{ODEtilde}) as $\widetilde{I}=C$, where $\widetilde{I}$ is defined
by replacing $y'$ (by the constant $k$) on $I$. By hypothesis, $I$ is elementary and,
consequently, $\widetilde{I}$ is also elementary. Therefore, by the
Theorem due to Prelle and Singer \cite{PS} and the Shtokhamer's extension \cite{Shtokhamer}, the integrating factor $\widetilde{R}$ (defined
by replacing $y'$ (by the constant $k$) on $R$) for equation
(\ref{ODEtilde}) can be written as $\prod_i
{\widetilde{f}_i}^{\,\,n_i}$, where $\widetilde{f}_i$ are irreducible
polynomials in $(x,y,S)$. We could proceed analogously for the other two
possible pairings of variables (i.e., $(x,y')$ and $(y,y')$). More
explicitly, in the reasoning leading to equation (\ref{PSbarra}), we
could have imposed the compatibility condition $I_{xy'}=I_{y'x}$
and, in the same fashion we did above for the pairing $(x,y)$, conclude
that the integrating factor can be written as a product of irreducible
polynomials in $(x,y',S)$. The same can be done for the pairing $(y,y')$.
So, we can conclude that $\overline{R}$ can be written as a
product of irreducible polynomials in $(x,y,y',S)$. Since
$\overline{R}\equiv(\frac{R}{N})$, and $N$ is a polynomial in
$(x,y,y')$, finally we have that $R = \prod_i f^{n_i}_i$,
where $f_i$ are irreducible polynomials in $(x,y,y',S)$ and $n_i$ are non-zero rational numbers.

Let us now prove that the $f_i$ are eigenpolynomials of the operator ${\cal D}$
or factors of $N$. From equation (\ref{SR1}) and remembering that
$\overline{R}\equiv(\frac{R}{N})$, we have:

\begin{equation}
\label{DRbarra}
\frac{D[\overline{R}\,N]}{\overline{R}\,N} = -S - \frac{N\,M_{y'}-M\,N_{y'}}{N^2}
\end{equation}
multiplying both sides of (\ref{DRbarra}) by $N^2$, we get
\begin{equation}
\label{DRbarra2}
N^2\,\frac{D[\overline{R}]}{\overline{R}}+N\,D[N] = - N^2\,S - (N\,M_{y'}-M\,N_{y'})
\end{equation}
and finally,
\begin{equation}
\label{DRbarra3}
N\,\frac{{\cal D}[\overline{R}]}{\overline{R}} = -N\,D[N]  - N^2\,S - (N\,M_{y'}-M\,N_{y'}).
\end{equation}
Since the right-hand side of (\ref{DRbarra3}) is a polynomial in
$(x,y,y',S)$, so is $N\,\frac{{\cal D}[\overline{R}]}{\overline{R}}$. Using
(\ref{Rfi}) this implies that
\begin{equation}
\label{DRRn}
N\,\frac{{\cal D}[\overline{R}]}{\overline{R}} = \sum_i\,n_i\,N\,{\frac{{\cal D}[f_i]}{f_i}}.
\end{equation}
is a polynomial. Since $f_i$ are irreducible, independent polynomials,
we may conclude that: either $f_i|{\cal D}[f_i]$ or $f_i$ is a factor of $N$.$\Box$
\bigskip

In order to obtain the integrating factor $R$, we are going to use
eq.(\ref{SR1}). Dividing it by $R$, we get:
\begin{equation}
\label{DRR}
D[R]/R  = - (S + \phi_{y'}) = - (S + \frac{NM_{y'}-MN_{y'}}{N^2})
\end{equation}

Multiplying both sides of the equation above by $N^2$, one has:
\begin{equation}
\label{DRR2}
N^2 D[R]/R  = - (SN^2 + NM_{y'}-MN_{y'})\,\,\,\Rightarrow\,\,\,N {\cal D}[R]/R  = - (SN^2 + NM_{y'}-MN_{y'})
\end{equation}

Due to the results of theorem 3, the equation (\ref{DRR2}) above
it can be solved in the same manner as in the methods inspired on
the Prelle- Singer Procedure. Once $R$ is found as a function of
$(x,y,y',S)$, we can substitute the known $S$ (see section
\ref{NewTheo}) and, finally, from eqs.(\ref{compatcond}), find the
first order invariant via quadratures:
$$I(x,y,y') = \int \!R\left (\phi+S{\it y'}\right ){dx} \,\,- \int \! \hbox{\large [} RS+{\frac {\partial}{\partial y}}\int
\!R\left (\phi+S{\it y'}\right ){dx} \hbox{\large ]} {dy} \,\,- $$
\begin{equation}
\label{C1}
\int \!\left[R+
{\frac {\partial }{\partial {\it y'}}}\left (\int \!R\left (\phi+S{
\it y'}\right ){dx}-\int \!\hbox{\large [}RS+{\frac {\partial }{\partial
y}}\int \!R
\left (\phi+S{\it y'}\right ){dx}\hbox{\large ]}{dy}\right
)\right]{d{\it y'}}.
\end{equation}
\noindent

That concludes the presentation of our proposed approach which is a semi-algorithm procedure to
reduce soodes of the form of eq.(\ref{soode}).

\subsection{Example}
\label{example}

Here we are going to briefly introduce an example of a
non-trivial\footnote{For instance, the MAPLE package, release 7, could not reduce it.}
SOODE that illustrates the results displayed above.

Consider the SOODE:

\begin{equation}
y'' = -\frac{\left( y'  \right) ^{2}+2\, \left( y'  \right) ^{3}-1-2\,y'
}{-x-(y')x-(y')y-1+(y')^2}
\end{equation}

For this SOODE, the operator $D$ becomes:
\begin{eqnarray}
D &=& \left(y'^{2}+2\,y'^{3}-1-2\,y'\right)\,\partial_{y'}+\left( x+y'\,x+y'\,y+1-y'^{2}  \right)\,y'\,\partial_{y}+ \nonumber \\
&&\left( x+y'\,x+y'\,y+1-y'^{2}  \right)\,\partial_{x},
\end{eqnarray}
and the operator $D_S$ is:
\begin{equation}
D_S=N\,D + {\cal M}\,\partial_s
\end{equation}
where ${\cal M}=2\,{s}^{2}xy'y+2\,{s}^{2}{y'}^{2}xy+7\,s{y'}^{2}x+4\,s{y'}^{3}x+{s}^{2}+{s
}^{2}{x}^{2}+2\,{s}^{2}x-2\,{s}^{2}{y'}^{2}+{s}^{2}{y'}^{4}-sx+4\,s{y'}^{
2}-2\,s{y'}^{4}+sy+4\,s{y'}^{3}y-2\,{s}^{2}{y'}^{2}x+{s}^{2}{y'}^{2}{y}^{2
}-2\,{s}^{2}{y'}^{3}y+2\,{s}^{2}y'{x}^{2}+2\,sy'x-2\,s-y'+{s}^{2}{y'}^{2}{x
}^{2}-2\,{s}^{2}{y'}^{3}x+2\,{s}^{2}y'y-2\,{y'}^{2}+{y'}^{3}+2\,{y'}^{4}+s{
y'}^{2}y+2\,{s}^{2}y'x$.

The polynomial $P= \left( x+y'x+y'y+1-{y'}^{2} \right)
S+1-{y'}^{2}$ is a eigenpolynomial of the $D_S$ operator. So,
$P=0$ defines $S$ as:
\begin{equation}
S=-{\frac {-1+{y'}^{2}}{x+y'x+y'y+1-{y'}^{2}}}
\end{equation}

In order to calculate the integrating factor $R$, we need (see
equations 31 e 32) to find the eigenpolynomials of the $D$
operator. These are found to be (up to the first degree):
\begin{eqnarray}
f_1 & = & 1+y'\nonumber\\
f_2 & = & 1-y'\nonumber\\
f_3 & = & 1+2\,y'\nonumber
\end{eqnarray}
and, for this particular case, these are sufficient (together with $N$) to build the
integrating factor $R$. Combining equations 31 and 32 and solving for the $n_i$'s we have that
$\{ n_1=-3/2, n_2=-3/2, n_3=0 \}$. So, $R$ is found to be\footnote{Note that $R$ is formed by the product $f_1^{n_1}\,f_2^{n_2}\,f_3^{n_3}\,N^{1}=f_1^{-3/2}\,f_2^{-
3/2}\,N^{1}$.}:

\begin{equation}
R = {\frac {x+y'x+y'y+1-{y'}^{2}}{ \left(  \left( y'-1 \right)  \left( 1+y'
 \right)  \right) ^{3/2}}}
\end{equation}

This leads to the first order invariant given by:
\begin{equation}
I = {\frac {x+y+y'x}{\sqrt {-1+{y'}^{2}}}}+\ln  \left( y'+\sqrt {-1+{y'}^{2}}
 \right).
\end{equation}

\section{Computational considerations}
\label{algol}

Here, we are going to introduce an alternative way of calculating
the integrating factor for a given SOODE. Why do we call this
section computational considerations? The reason is that the
material contained in this section is not applicable to all
SOODEs\footnote{Note that the method presented on the previous
section is a general semi-algorithmic approach to deal with SOODEs
of the form (\ref{soode}) that present an elementary first order
invariant.} of the form of eq.(\ref{soode}) and, indeed, we do not
have a criteria of applicability. But, on the other hand, the
method we are about to present is ``less costly'' computationally,
therefore could be useful in practical applications. In the
following, we will be using some results from the Lie theory.

Let us show a relation between the function $S$ and a symmetry
of the SOODE (\ref{soode}). Making the following transformation

\begin{equation}
\label{Seta}
S = - \frac{D[\eta]}{\eta}
\end{equation}
eq.(\ref{DS}) becomes

\begin{equation}
\label{eqeta}
D^2[\eta] = \phi_{y'}\,D[\eta] + \phi_{y}\,\eta.
\end{equation}

From Lie theory we can see that eq.(\ref{eqeta}) represents the
condition for a SOODE (\ref{soode}) to have a symmetry $[0,\eta]$. So, from
(\ref{Seta}) we can find a symmetry given by

\begin{equation}
\label{etaS}
\eta = e^{-\int S\,dx}.
\end{equation}

Looking at eq.(\ref{di}) we can infer that $R$ is also an integrating
factor for the auxiliary FOODE defined by
\begin{equation}
\label{auxi}
\frac{dy'}{dy} = S,
\end{equation}
where $x$ is regarded as a parameter. Besides, for this FOODE,
$[\eta,D[\eta]]$ is a point symmetry. So, $R$ is given by
\begin{equation}
\label{RSeta}
R = \frac{1}{\eta\,S+D[\eta]}.
\end{equation}

If in (\ref{RSeta}) we use $\eta$ defined by (\ref{etaS}) we would get a
singular $R$. However, eq.(\ref{eqeta}) has another solution
independent from $\eta$ given by

\begin{equation}
\label{eta2}
\overline{\eta} = \eta \int \frac{e^{\int \phi_{y'}dx}}{\eta^2} \,dx.
\end{equation}
Using this $\overline{\eta}$ in (\ref{RSeta}) we get $R$ and, once this
is done, we can calculate the first integral $I$ by using
simple quadratures.

Actually, trying to solve eq.(\ref{etaS}) could be complex. The
operator $\int S\,dx$ that appears on equation (\ref{etaS}) is
meant to be the inverse of operator $D$ (defined on
(\ref{DS}))\footnote{Actually, this meaning for the operator
$\int$ is true for the whole of the present section.}. As one
might expect, actually finding $\eta$ from (\ref{etaS}) could be
trick. Actually, in general, it will be impossible to integrate
(\ref{etaS}) and that is the reason why this shortcut is not
applicable in general.

\subsection{Example}

Consider the SOODE

\begin{equation}
\label{eqex}
y''=-\,{\frac {2\,{y'}^{2}{x}^{2}-2\,{y'}^{2}-2\,xy'y-{y}^{2}{x}^{4}+2\,{y
}^{2}{x}^{2}-{y}^{2}}{2 y \left( {x}^{2}-1 \right) }}.
\end{equation}
Constructing the $D_S$ operator we get that
\begin{equation}
\label{Dsex}
{y}^{2} \left( {x}^{2}-1 \right) -{S}^{2}{y}^{2}-{y'}^{2}+2\,Sy'y
\end{equation}
is an eigenpolynomial of it. Then $S$ is given by $S = y'/y + \sqrt{x^2-
1}$ and $\eta$, $\overline{\eta}$ are respectively
\begin{equation}
\label{etaetab}
\eta = {\frac {\sqrt {x+\sqrt {{x}^{2}-1}}}{y\,{e^{\frac{x\sqrt {{x}^{2}-1}}{2}}}}}\,, \,\,\,\,\,
\overline{\eta} = \frac{\sqrt {x+\sqrt {{x}^{2}-1}}\int \!{
\frac {\sqrt {x-1}\sqrt {x+1}{e^{x\sqrt {{x}^{2}-1}}}}{x+\sqrt {{x}^{2
}-1}}}{dx}}{y\,{e^{\frac{x\sqrt {{x}^{2}-1}}{2}}}}.
\end{equation}

$R$ can be written as
\begin{equation}
\label{Rex}
R = {\frac {y\sqrt {x+\sqrt {{x}^{2}-1}}}{
\sqrt {x^2-1}{e^{\frac{x\sqrt {{x}^{2}-1}}{2}}}}}
\end{equation}
leading to the following first integral:
\begin{equation}
\label{Iex}
I = {\frac {\sqrt {x+\sqrt {{x}^{2}-1}}
 \left( 2y'y+\sqrt {{x}^{2}-1}{y}^{2} \right) }
{\sqrt {x^2-1}e^{\frac{x\sqrt {{x}^{2}-1}}{2}}}}.
\end{equation}

\section{Conclusion}
\label{conclu}

In \cite{PS2}, we have developed a method, based on a conjecture,
to deal with SOODEs that presented an elementary solution (possessing
two elementary first order invariants).

In that same paper, we have introduced a function $S$ to transform the
Pfaffian equation related to the particular SOODE under consideration
into a 1-form proportional to differential of the first order
differential invariant. That function $S$ was instrumental in finding
the integrating factor for the SOODE.

Here, in the present paper, we introduce many theoretical results
concerning that function $S$ and present an way to calculated it via a
Darboux-type procedure. Here also the function $S$ is used to produce
the integrating factor.

We then demonstrate general results about the structure of the
integrating factor $R$ and that we can calculate it
using a procedure very similar to the one inspired by the original work
by Prelle-Singer \cite{PS} (applicable for FOODEs). These results assure that we have a semi-
algorithmic procedure to deal with rational SOODEs, presenting at least
one elementary first order invariant, i.e., there is no need to posses
two such invariants and, consequently, we can cover a much broader class
of SOODEs than before \cite{PS2}.

In the above section, we introduce an alternative way to calculate
$R$ from the knowledge of $S$ for restricted cases. The motivation
for that method is that, the general method, sometimes, can be
computationally very demanding and it may be worth to have a go in
the ``shortcut'' before embarking on long calculations (despite
the general case being the sure think).

We are searching for better algorithms for the many steps of the
method presented here in order to make it computationally more
efficient. We are also working on a full computational
implementation of the method as it stands today.


\end{document}